\documentclass[apj]{emulateapj}
\usepackage{epsfig}

\newcommand{\dm}{$\Delta m_{12}$}
\newcommand{\dmt}{$\Delta m_{12} \geq 2$}
\newcommand{\dmfo}{$\Delta m_{14}$}
\newcommand{\dmtfo}{$\Delta m_{14} \geq 2.5$}

\newcommand{\Mr}{$M_{r}$}

\shorttitle{A statistical study of the luminosity gap in galaxy
groups} \shortauthors{Tavasoli et al} \slugcomment{}

\begin{document}
\title{A statistical study of the luminosity gap in
galaxy groups}

\author{Saeed Tavasoli  }
\affil{Department of Physics, Ferdowsi University of Mashhad,
Mashhad, Iran} \affil{School of Astronomy and Astrophysics,
Institute for Research in Fundamental Sciences (IPM), P. O. Box
19395-5531, Tehran, Iran} \email{tavasoli@ipm.ir}

\author{Habib G.~Khosroshahi}
\affil{School of Astronomy and Astrophysics, Institute for Research
in Fundamental Sciences (IPM), P. O. Box 19395-5531, Tehran, Iran}

\author{Ali~Koohpaee}
\affil{School of Astronomy and Astrophysics, Institute for Research
in Fundamental Sciences (IPM), P. O. Box 19395-5531, Tehran, Iran}
\affil{Faculty of Science, Khajeh Nasir Toosi University of
Technology, Tehran, Iran}

\author{Hadi~Rahmani}
\affil{Inter-University Center for Astronomy and Astrophysics,
Post-Bag 4, Ganeshkhind, Pune 411007, India}

 \and

\author{Jamshid~Ghanbari}
\affil{Department of Physics, Ferdowsi University of Mashhad,
Mashhad, Iran}

\begin{abstract}
The luminosity gap between the two brightest members of galaxy
groups and clusters is thought to offer a strong test for the models
of galaxy formation and evolution. This study focuses on the
statistics of the luminosity gap in galaxy groups, in particular
fossil groups, e.g. large luminosity gap, in an analogy with the
same in a cosmological simulation. We use spectroscopic legacy data
of seventh data release (DR7) of \emph{SDSS}, to extract a volume
limited sample of galaxy groups utilizing modified
friends-of-friends (mFoF) algorithm. Attention is paid to galaxy
groups with the brightest group galaxy (BGG) more luminous than \Mr
= -22. An initial sample of 620 groups in which 109 optical fossil
groups, where the luminosity gap exceeds 2 magnitude, were
identified. We compare the statistics of the luminosity gap in
galaxy groups at low mass range from the SDSS with the same in the
Millennium simulations where galaxies are modeled semi-analytically.
We show that the BGGs residing in galaxy groups with large
luminosity gap, i.e. fossil groups, are on average brighter and live
in lower mass halos with respect to their counter parts in
non-fossil systems. Although low mass galaxy groups are thought to
have recently formed, we show that in galaxy groups with 15 galaxies
brighter than $M_r\ge -19.5$, evolutionary process are most likely
to be responsible for the large luminosity gap. We also examine a
new probe of finding fossil group, \dmtfo, and find that the fossil
group selected according to new probe are more abundant than those
selected using the conventional probe, \dmt, in low halo mass
regime, $\leq 10^{14} M_{\odot} $ . In addition we extend the
recently introduced observational probe based on the luminosity gap,
the butterfly diagram, to galaxy groups and study the probe as a
function of halo mass. This probe can, in conjunction with the
luminosity function, help to fine tune the semi-analytic models of
galaxies employed in the cosmological simulations.

\end{abstract}


\section{Introduction}

Galaxy groups are key systems in advancing our understanding of
structure formation and evolution in the universe as, in a
hierarchical framework, they span the regime between individual
galaxies and massive clusters. Advances in the cold dark matter
cosmological simulations and improvements on the particle mass
resolution has provided the opportunity to study galaxy groups or low
mass halos. The Millennium simulation \citep{spr05} is a recent
example. The semi-analytic models of galaxy formation are also
employed to understand the formation and evolution of galaxies in
cosmological context.

The key factor in the success of this attempt is the availability of
observational constraints to help fine tune the models. While galaxy
luminosity function has been widely used as a strong constraint, the
number of physical processes and the free parameters in the models
limits the achievements.

There is a class of galaxy groups, dubbed as fossil groups
\citep{p94}, which are the archetypal relaxed systems and arguably
the end product of the galaxy mergers within the group. The
selection criteria for fossils is outlined by \cite{J03}, i.e.
groups with an X-ray luminosity of $L_{x,bol} \ge 0.25 \times
10^{42}$ h$^{-2}$ergs$^{-1}$, and a minimum luminosity difference of
2 magnitude between the first- and second-ranked galaxies ($\Delta
M_{12}\ge 2$), within half the projected $R_{200}$ radius, where
$R_{200}$ is the radius within which the mean density is 200 times
the critical density of the Universe.

Implementation of cosmological simulations in which the evolution of
the halos can be traced, has given strong evidences that fossil
groups, or galaxy systems with large luminosity gap $\Delta M_{12}$,
on average form earlier than galaxy groups with small luminosity gap
(\citealt{dar07,dar10}). The early formation epoch for fossil groups
had been argued previously based on observations of about a dozen
such systems. They included the study of X-ray scaling relations
\citep{kh07} as well as the morphological studies of the brightest
group galaxies (BGGs) in fossil groups \citep{kh06}.

A statistical comparison between the properties of observed fossil
groups and those in the cosmological simulations requires dozens of
such systems to be identified observationally which is currently
unavailable due to the nature of existing X-ray surveys (often
shallow or very limited in angular coverage).  \cite{san07} have
cross-correlated optical and the Sloan Digital Sky Survey (SDSS) and
the ROSAT All Sky Survey (RASS) (Voges et al. 1999) and identified
34 fossil group candidates covering a wide range of redshift. The
applied methodology, however, does not produce a complete sample of
fossil groups.

Although conventionally an X-ray threshold has been applied to the
IGM in fossil groups a lot can be learned from the selection based
on optical criterion alone, e.g. ``optical fossils'' \citep{dar07}.
Identifying low-mass fossil systems is hampered by the fact that
groups are under-represented in existing X-ray catalogs, however,
based on simulation data \cite{dar07} have shown that even applying
X-ray criteria, the fraction of late-formed systems that are
spuriously identified as fossils is $\sim 4-8$ per cent, almost
independent of halo mass.

The main driver of this study is to find out the extend at which the
luminosity gap in low mass groups can shed light on the formation of
these systems and quantify the observed properties related to the
luminosity gap which can be compared with semi-analytic models.  As
the large luminosity gap between the galaxies in a galaxy group can
have a statistical origin, it would be therefore useful to find out
to what extend the large luminosity gap is a representative of the
evolutionary processes as opposed to have been originated
statistically. Furthermore we provide observational measures which
can be used to constrain the semi-analytic models of galaxy
formation.

The paper is organized as follows; Section 2 describes the databases
and the simulations. Algorithm used to extract the sample of groups
in observation is described in section 3. Biases and completeness of
the sample is discussed in section 4. In section 5 we present the
derived group properties. Selection criteria for identifying fossil
group is discussed in section 6. Results are presented and discussed
in sections 7 to 9 with concluding remarks presented in section 10.

For this study we assume, $h=0.7$, $\Omega_{m}=0.3$, and
$\Omega_{\Lambda}=0.7$.

\section{Data}
\subsection{Observation}

We use the legacy archive of the latest data release of the SDSS,
DR7 \citep {aba09} which covers 8,423 square degrees in imaging data
(which contains roughly 360 million distinct photometric objects)
and the spectroscopic survey mapped 8032 square degrees (1,640,000
objects with measured spectra). We confine the study to those
objects identified in SDSS as galaxies. In this study, we use both
the photometric and the spectroscopic data.

Initially we restrict the study to galaxies within the redshift
range of $0.02 < z < 0.17$, retaining only those with clean spectra
(flag=0) providing us with a sample of 601,981 galaxies. The lower
limit on the redshift is dictated recalling that, objects with $z <
0.02$ are often stars and the upper limit on redshift is placed in
order to confine the study to the range where the spectroscopic
sample is complete. It worths stressing that, the main galaxy
spectroscopic sample of SDSS is restricted to Petrosian magnitude,
$r_{petro} < 17.77$.

We extract objects with $r_{petro} < 21.0$, identified as galaxy,
from the photometric sample of SDSS DR7. For galaxies fainter than
$r_{petro} = 21.0$, the error in apparent magnitude increases
rapidly (\citealt{oya08}). For objects satisfying the above
conditions, we retrieved the Petrosian magnitudes (\citealt{pet76};
\citealt{str02}) and K-corrected them using the method described in
\cite{bla03a}.

\subsection{Simulation}

In this paper we use the galaxy group catalog from the Millennium
simulation \citep{spr05} and the associated semi-analytic model of
galaxy formation by \cite{bow06}. Below we briefly describe these
simulations.

The Millennium simulation utilizes a $\Lambda CDM$ cosmological
model to follow structure formation from $z=127$ up to the present
epoch. Starting with an inflationary, dark-matter dominated
universe, structures form through the bottom-up hierarchy, i.e.,
starting with small scale density fluctuations resulting in large
scale structures we observe today. Evolution of $2160^{3}$ particles
each with $8.6 \times 10^{8} h^{-1} M_{\odot} $ has been followed
from $z=127$ up to present day ($64$ time-slices of positions and
velocities were stored, separated logarithmically between $z=127$
and $z=0$) through a comoving box of $ 500 Mpc $ on each side,
\citep{spr05} with $\Omega_{\Lambda} = 0.75$, $\Omega_{M} = 0.25$,
$\Omega_{b} = 0.045$, $h = 0.73$, $n = 1$ and $\sigma_{8} = 0.9$
based on WMAP observations, \citep{spe03} and 2dF galaxy redshift
survey, \citep{col01}. The evolution of a structure has been
followed only if it composed of at least, 20 particles (equivalent
mass is $1.72 \times 10^{10} h^{-1} M_{\odot}$) at that epoch
\citep{spr05}. In addition, friends-of-friends algorithm was
exploited to extract haloes with densities at least $200$ times the
critical density and substructures were identified using SUBFIND
algorithm developed by \cite{spr01}.

The underlying dark matter haloes of the Millennium simulation was
used to simulate the growth of galaxies, by self-consistently
implementing a semi-analytic model of galaxies on the outputs of the
Millennium simulation \citep{bow06}.  This semi-analytic model
accounts for the feedback from supernova explosions as well as AGNs
in modeling the massive halos. In addition, AGN feedback has been
considered as an operation which quenches the star-formation and
star-bursts are triggered both by merging and disk-instabilities.
\cite{bow06} predicts the luminosity function of galaxies in B and K
band for the present epoch as well as the present day color
distribution. For the purpose of this study, we use the catalogue of
\cite{bow06} semi-analytic model and retrieved the absolute r-band
magnitude for group members, the halo mass of the group as well as
the mass of the subhalo containing the BGG at the present epoch.

\subsection{Monte Carlo Simulation}

The idea behind performing the Monte-Carlo simulation is that the
luminosity gap between the two brightest members (or between any
other members) of a galaxy system can also have a statistical
origin.

We expect large luminosity gaps to appear preferentially in groups
with few members. In order to understand the extend at which the
luminosity gap with statistical origin affects our analysis and
results, we perform a Monte-Carlo simulation by drawing random
samples from an underlying Schechter function (\citealt{sch76}).

We generate a sample of $\sim 10^{6}$ galaxy groups with different
number of members within the completeness of the observed sample
described in section 4. We choose the underlying luminosity function
to have Schechter parameters of $\alpha = -1.08$ and $M_{r}^{*} =
-21.62$ as in \cite{zan06}, with galaxy luminosity ranging from
$M_r=-16$ to $-26$. The chosen values are the best fit Luminosity
function for groups based on the r-band SDSS data.

\section{Selecting groups in observational data}
\subsection{The Algorithm}

To identify galaxy groups, we use the modified friends-of-friends
(mFoF) algorithm developed by \cite{par95} which is a modification
of FoF \citep{huch82}, one of the most frequently applied methods
for finding structures in redshift surveys. The starting point in
this algorithm is that every galaxy can be the center of a group of
galaxies. The algorithm, however, will gradually lead to the most
probable center, for a given search radius.

Given this, we start by looking for companions of each galaxy within
a specified search physical radius, $r_{s}$. A galaxy is companion
to the chosen galaxy if it falls within the search radius in the
plane of the sky and within the redshift interval of $\Delta z$. For
this study we choose $r_s$ to be the equivalent of 500 kpc at the
redshift of the galaxy in the spectroscopic sample and $\Delta
z$=0.002.

For the first generation of the groups, we count the number of
companions with the above constraints for each galaxy in the
spectroscopic sample described in section 2.1. Naturally most of the
members will be shared by galaxy groups found in the first
generation. Comparing overlapping groups, those with fewer number of
member will be removed from the list if they fall within the $r_s$
measured from the center of the richer group and $\Delta z$=0.002.
This results in the second generation of the galaxy groups. Having
this, we calculate the mean redshift and luminosity weighted center
of each of the surviving groups and the process of removing the
small groups overlapping with larger groups continues until there is
no common members between any two groups of galaxies.

As it was described, mFoF works with two free parameters,$r_s$ and
$\Delta z$, and their values are set depending on the science goals
of the study, e.g. \cite{mil05} and \cite{vonder07} have been used
$r_s=1 Mpc$ ; $\Delta z$=0.002 and $r_s=2 Mpc$ ; $\Delta z$=0.01 to
identify cluster of galaxies, respectively. Also, \cite{ein05}
adopted $r_s=0.5 Mpc$ ; $\Delta z$=0.002 to make catalogue of
group/cluster. Here we aim at studying galaxy groups and low mass
haloes for which we find our choice suitable in comparison to the
existing galaxy group catalogs. We extracted 81614 and 6956 groups
with more than 2 and 5 spectroscopic members $(richness)$,
respectively.

There exist several catalogues of groups and clusters of galaxies
extracted from the SDSS data: Catalogues by Goto (2005, DR2), Miller
et al. (2005, DR2), Merchan  and Zandivarez (2005, DR3), Berlind et
al.(2006, DR3), Zandivarez et al. (2006, DR4), Yoon et al. (2008,
DR5) and Tago et al. (2008, DR5). In each of the above mentioned
catalogs, criteria implemented through the group-finder algorithm
has been chosen on the basis of the type of the corresponding study.
To check if our extracted groups, based on conditions adopted here,
overlap with those from previous studies, we performed a
cross-matching between our extracted catalog and two other popular
group samples in the literature: the Abell cluster (\citealt{ab58})
and C4 (\citealt{mil05}) catalogs. It worth stressing that although
C4 is known as a ``galaxy cluster catalog'', it has a wide mass
range making it suitable for the comparison with our ``galaxy
group'' catalog.

For the redshift of Abell clusters we used published data by
\cite{stru87} in which there are 838 with determined redshifts out
of 2712 Abell clusters. There are 318 Abell clusters that lie in the
area of sky coverage for the SDSS DR7 and have redshifts in the
range of $0.02 < z < 0.17$. For our groups of richness $\ge 5$
(spectroscopic members), 195 out of 318 Abell clusters $(61\%)$ are
matched within $10''$ of our group center. In addition, we selected
the groups of richness $\ge2$ and in this case, 283 out of 318 Abell
clusters are matched ($88\%$). The C4 catalog was generated using a
cluster finding algorithm which identifies clusters as over
densities in a seven dimensional space of position and colors
(\citealt{mil05}). The catalog contains 748 clusters with richness
more than 10 members brighter than r = 17.7. Applying the above
search criteria we found 549 of 748 C4 clusters, $73\%$ within $0.02
< z < 0.17$. Reducing the number of galaxies per group, $\ge2$, we
would find $96\%$ of C4 clusters, which shows a very large overlap
between our extracted groups and the C4 catalog.

\begin{figure}
\centering
\includegraphics[width=80mm]{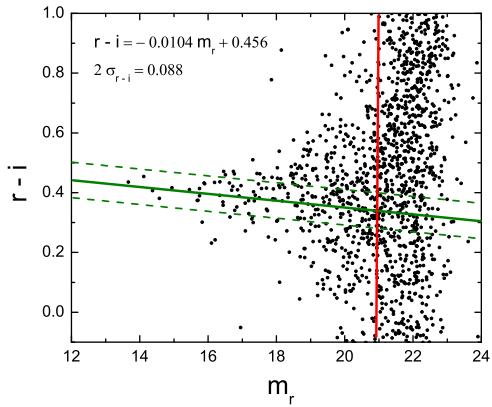}
\caption{Color-magnitude relation for members residing within a
centered at the BGG position with a radius of 0.5 Mpc at the BGG
redshift. The 2$\sigma$ boarders and restrictions on apparent
magnitude are also illustrated. \label{fig1}}
\end{figure}

\subsection{Color-Magnitude Relation}

As noted earlier, the main spectroscopic sample of the SDSS is
complete to the Petrosian $r < 17.77$. However, the photometric
sample reaches lower luminosities. Moreover, the tiling algorithm
used in SDSS spectroscopic survey \citep{bla03b}, leaves some
galaxies unobserved spectroscopically because of fiber collisions.
\cite{yoon08} estimated the spectroscopic completeness of the SDSS
DR5 to be $f_{spec}\sim65\%$ for rich clusters. In order to have a
realistic estimate of the optical luminosity and the richness of the
extracted groups we use the color magnitude relation (hereafter CMR)
to identify additional group member candidates. Observational
evidences have shown that the bulk of the early-type galaxies in
clusters and groups lie along a linear CMR. This relation is shown
to have a small scatter (e.g., \citealt{bow92,peb02,koda98}) and can
be used as a robust method for finding galaxy systems (e.g.,
\citealt{gla00}). Although there are uncertainties in the CMR method
of group membership identification, we use the photometric members
only to estimate the global properties, such as the number of
galaxies per groups and the total luminosity of the group.

In order to incorporate the photometric data into the analysis, we
consider all the photometric and spectroscopic galaxies within an
angular distance equivalent to 500 kpc around the group center if
they fall on the CMR. It should be noted that, group center is
defined as the brightest member of the group. For this purpose, we
fit a line to the color-magnitude diagram of the spectroscopic
members if the group has at least three members with $M_{r} \le
-19.0$. We determine the slope and intercept of the CMR for each
group. We consider only those photometric ($m_{r} \le 21$) and
spectroscopic galaxies which fall within $\pm2\sigma $ of the color
magnitude relation defined by spectroscopic members. An example of a
fitted CMR to color magnitude diagram for one of our groups is shown
in Fig. \ref{fig1}.

Member selection using the color-magnitude relation method is
hampered by miss-identification of forground and background galaxies
as group members. In principle, miss-identification of members will
affect the total luminosity and in turn the estimated mass (\S 5.2).
In order to quantify the uncertainty in halo mass estimate we
randomly selected field regions in the SDSS. We chose galaxies
within a radius comparable to that of groups in each field region
and applied color-magnitude selection. The contribution of the field
galaxies on the halo mass of groups was found to be negligible.

We assume all the photometric group members found using the CMR to
have the mean redshift of the group defined by the spectroscopic
members.

\begin{figure}
\centering
\includegraphics[width=80mm]{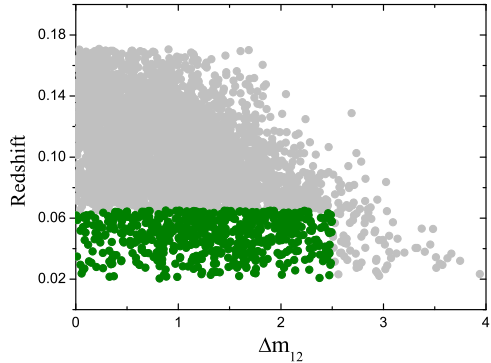}
\caption{4113 groups with $M_{r,BGG} \le -22$ are shown in redshift
vs. \dm $ $ diagram. The green rectangle area represents the
selected region which is free of biases.} \label{fig2}
\end{figure}

\section{Bias Correction and Completeness}

As noted in \S 2.1, we have used the main spectroscopic sample and
photometric legacy survey of SDSS DR7 as our spectroscopic and
photometric input. Statistical properties of the input data is
discussed here along with the constrains resulted from the
completeness and bias removal from the samples. For the photometric
sample, \cite{aba09} demonstrated that the sample is $95\%$ (95\%
completeness limit for point sources) complete to $r \approx 22.2$.
Whereas, the main spectroscopic sample of SDSS DR7 is complete to a
r-band petrosian magnitude \citep{pet76} of $r \simeq 17.77$,
corrected for Galactic extinction (\citealt{schl98,str02,aba09}).

The limiting magnitude of $r\simeq17.77$ is equal to $M_r\simeq -17$
at $z=0.02$ and $M_r\simeq -21.8$ at $z=0.17$. As we are interested
in large luminosity gap, i.e. \dmt$ $ magnitude, the above
completeness limit will naturally bias our analysis toward low
luminosity gap at higher end of the redshift distribution. This is
shown in Fig.\ref{fig2}.

To overcome this, we apply the \Mr $\le -22$ constraint to the BGGs
both in the observed SDSS groups (4113 groups) and in the
simulations (18843 groups). The constrain on the BGG luminosity is
also consistent with the value adapted for the $M^{*}$ parameter in
Schechter luminosity function in r-band. In addition this selection
helps to compensate the absence of X-ray data, given the correlation
of the BGG luminosity with the IGM X-ray emission
\citep{kh07,ellis06}, and thus reduces contamination from individual
galaxies that are not associated with a group scale halo.

To ensure that a luminosity gap up to 2.5 mag can be observed
without a bias, the second brightest galaxy should be at least as
luminous as -19.5 mag which corresponds to the limiting apparent
magnitude of the spectroscopic sample, i.e $r\simeq17.77$ mag at
$z=0.065$. In other words, our luminosity gap estimates less than
2.5 magnitude should be bias free for groups within $0.02\le z\le
0.065$. We therefore restrict the observed galaxy sample to the
above redshift range.

Applying the above mentioned restrictions (\dm $\leq 2.5$ and $z
\leq 0.065$) resulted in 109 fossils out of 620 groups (see. \S 6).

\section{Estimating Physical Parameters for Groups}

Based on available parameters for members of each galaxy group, we
have derived a number of quantities both for simulated and observed
groups which is discussed here.

\subsection{Velocity Dispersion and Virial Radius}

According to \cite{bee90}, a reliable method for measuring the
velocity dispersion in groups with few members is the so-called
\textit{gapper} estimator (e.g. \citealt{yang05}). Since our groups
catalogue contains low member groups we use this method to estimate
the line-of-sight velocity dispersion of each individual group.

The method involves ordering the set of recession velocities
${v_{i}}$ of the $N$ group members and defining gaps as

\begin{equation}
g_{i} = v_{i+1}-v_{i}, \quad\quad    i=1,2,...,N-1.
\end{equation}

The rest-frame velocity dispersion of our groups is then given by

\begin{equation}
\sigma_{gap} =\frac{\sqrt{\pi}}{(1+z_{group})N(N-1)}\sum_{i=1}^{N-1}
{w_{i}g_{i}}.
\end{equation}

where the weight is defined as $w_{i}=i(N-i)$ . As the central
galaxy in each group, is assumed to be at rest with respect to the
dark matter halo, the estimated velocity dispersion has to be
corrected by a factor of $\sqrt{N/(N-1)}$ \citep{yang05}. This
results in a final velocity dispersion given by

\begin{equation}
\sigma_{v} =\sqrt{\frac{N}{N-1}}\sigma_{gap}.
\end{equation}

As we will see in the next section, an estimation of the virial
radius of the galaxy groups is also required. Using the Virial
theorem, \cite{gir98} derives an approximation for the virial radius
of a spherical system as:

\begin{equation}
R_{vir} = \frac{\sqrt{3}}{10} \frac{\sigma_{v}}{H_{0}} \quad Mpc
\end{equation}

where $\sigma_{v}$ is the velocity dispersion of group members.

On the other hand, for the simulated groups, given that the halo
mass is derived directly by adding the mass of the dark matter
particles, $R_{vir}$ is calculated as follows:

\begin{equation}
R_{vir}= \frac{G M_{halo}}{\sigma_{v}^{2}} \quad  Mpc
\end{equation}

where $M_{halo}$ is the total mass of the halo and $\sigma_{v}$
denotes the velocity dispersion of group members.

\begin{figure}
\centering
\includegraphics[width=80mm]{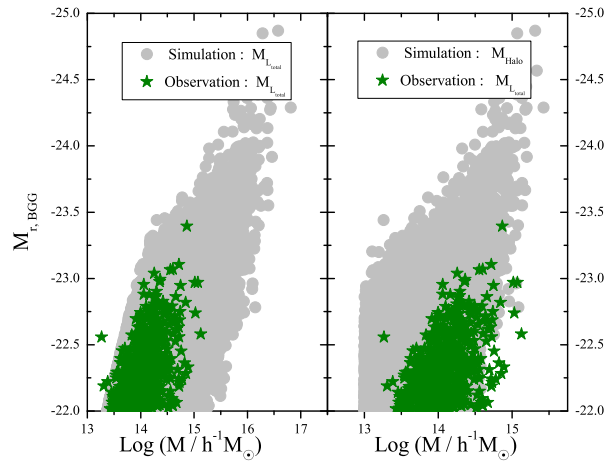} \caption{The distribution of the
galaxy groups in the plane of total mass of the group and the BGG
luminoisty. Total mass of the simulated groups are estimated based
on total luminosity Eq 6 (left) and dark matter particles (right).
The mass of observed groups are based on the total luminosity. It is
noticeable that, total mass obtained through total luminosity is in
a better agreement with the observations.} \label{fig3}
\end{figure}

\subsection{Halo Mass}

For observed groups, the total luminosity was used as diagnostic of
halo mass, as appeared to be a better mass estimator than the
line-of-sight velocity dispersion (\citealt{pop05,mil05,lin04}) :

\begin{equation}
log(h^{-1}M) \approx -2.46+1.45 log (h^{-2}L_{r})
\end{equation}

where $L_{r}$ is the total r-band luminosity of the galaxy group,
defined as the sum of the r-band luminosity of member galaxies with
$M_{r} < -19.5$, (\citealt{milo06}).

In addition to the measurement of $M_{halo}$ in the simulated
groups, we also estimate the halo mass using the empirical relation
(Eq. 6) for consistency. In Fig. \ref{fig3}, the two methods of mass
estimation are compared. Left panel compares the distribution of the
halos in the plane of total gravitational mass based on Eq. 6 and
the luminosity of the BGG for both the simulated and the observed
systems. The same is shown in the right panel with the mass of the
simulated halos obtained from the dark matter only. As expected, a
better consistency is achieved when the luminosity based estimation
is applied to the observations and the simulations. Therefore this
mass indicator is used in the our analysis to avoid any systematics.
The same method was applied in assigning a mass to the halos in the
Monte Carlo simulations (see. \S 2.3).

\begin{figure}
\centering
\includegraphics[width=80mm]{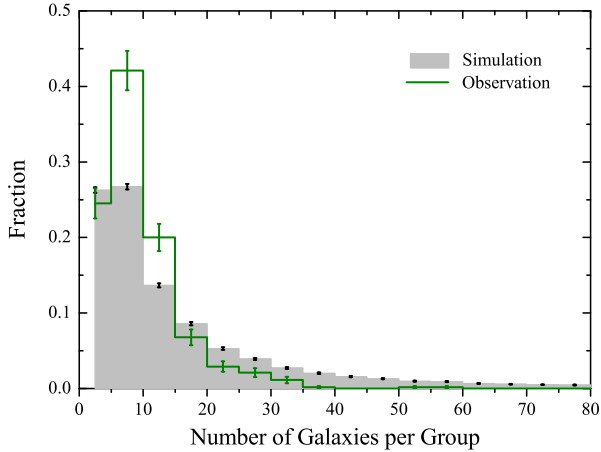}
\caption{The normalized histogram of number of galaxies per group is
compared between simulated and observed groups. The K-S test shows
that the probability of two samples selected from a common
distribution is $\sim$ $10^{-21}$.} \label{fig4}
\end{figure}

\subsection{Richness}

To obtain the number of galaxies per group we applied $M_{r} \leq
-19.5$ criterion to the observed and the simulated groups
(Millennium and Monte-carlo). The distribution of richness for the
observed and the simulated groups is shown in Fig. \ref{fig4}. This
plot illustrates that, richness of simulated and observed groups are
different which might be related to different group selection
procedures in simulation and observation data. The group selection
in simulations is based on a full knowledge of the three dimension
distribution of dark matter subhalos (e.g. galaxies), while as noted
earlier, the group selection in observations is based on two
dimension distribution of galaxies complemented by slices the
redshift space. However, the difference in shape of the
distributions has no impact on our conclusions.

In order to study the impact of biases on statistics of number of
galaxies per group, generated groups through the Monte Carlo
simulation had to span a large range of richnesses ($10^{4}$
simulations were carried out for each richness class of group).

\section{Selection of Fossil Groups}

In low-density environments, the merging of compact groups can lead
to the formation of the fossil groups (\citealt{vikh99,J03}). These
class of systems have masses which are comparable to those of normal
groups and clusters of galaxies. Observationally, the identification
of fossils have so far been based on the definition of fossil group
suggested by \cite{J03}. In \cite{J03} definition, a fossil group
has an X-ray luminosity of $L_{X,bol} \geq 0.25$ $\times$ $10 ^{42}$
$h^{-2}$ $erg$ $ s^{-1}$, and the dominant galaxy is at least 2
magnitudes brighter (in r-band) than the second ranked galaxy within
$0.5 R_{200}$ of the group. Assuming a circular orbit, the
$L^{\ast}$ galaxies merge with the central galaxy within a few Gyr
due to orbital decay as a result of the dynamical friction
(\citealt{J03}).

The existing X-ray surveys overlapping the area covered by the SDSS
survey are either wide and shallow (ROSAT All Sky Survey,
\citealt{vog99}) or very small and deep. As a result, facing the
lack of suitable X-ray data we limit the study to optical fossils or
galaxy groups with a large luminosity gap (\citealt{dar07}). The
X-ray criterion is conventionally used to ensure that only
group-mass halos have been selected. As noted earlier we compensate
this by the introduction of lower limit in the BGG luminosity.

The luminosity gap was measured within half a virial radius,
$R_{vir}$, derived using equations 4 and 5. This is slightly
different to the definition of the \cite{J03}, however the
difference applies to both the simulations and observations,
consistently.

Similar to observational constrains discussed in \S 4 ,the same
conditions was applied to simulation data to select optical fossil
groups with the exception of the X-ray luminosity criterion. Also
following \cite{dar07}, we limit the analysis to groups with
$log(M_{halo}/h^{-1} M_{\odot})> 13$.

These conditions resulted in 1993 fossil systems out of 16688
groups, or roughly $12\%$. Furthermore, fossil groups through the
Monte Carlo simulation, were defined as groups $\Delta m_{12} \geq
2$ and $M_{r,BGG} \leq -22$ in accordance with definitions for
observation and simulation data.

In order to compare the properties of fossil with non-fossil groups,
we introduced control groups with $\Delta m_{12} \leq 0.5$ and
satisfying the remaining conditions mentioned above (e.g. $M_{r,BGG}
\leq -22$), in observations and simulations. This provides us with
85 and 2419 control groups in observed and simulated catalogs,
respectively.

\begin{figure}
\centering
\includegraphics[angle=0,width=80mm, clip=true]{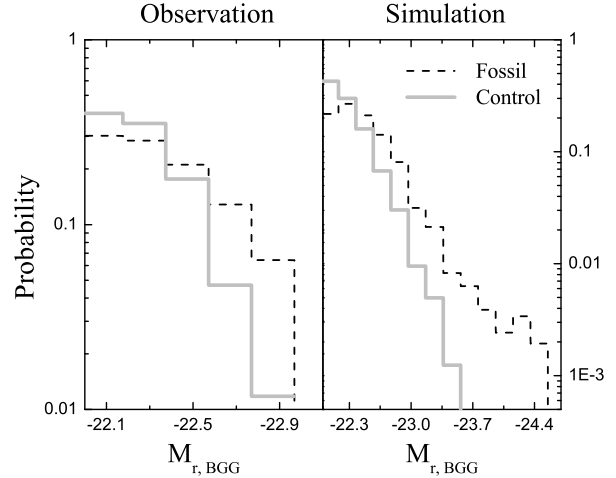}
\caption{The distribution of r-band absolute magnitude of fossil
BGGs is compared to the same in control groups for observational and
simulation samples. Fossil BGGs are brighter than control BGGs. The
K-S test shows that the probability of control and fossil samples
being drawn from similar distribution are $1.2\times10^{-3}$ and
$3.6\times10^{-64}$ in observation and simulation samples,
respectively.} \label{fig5}
\end{figure}

\begin{figure}
\centering
\includegraphics[angle=0,width=80mm, clip=true]{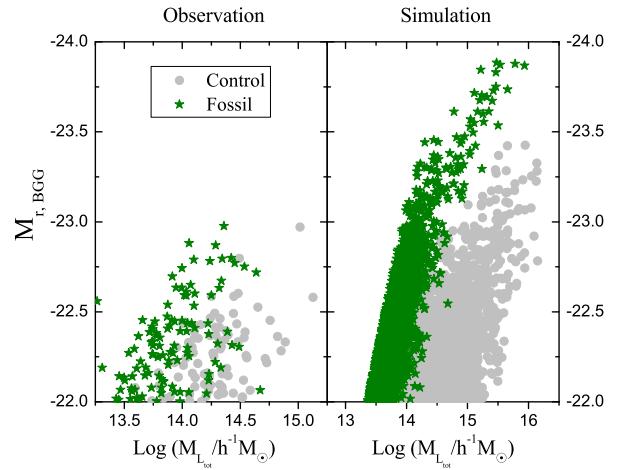}
\caption{The distribution of BGG absolute magnitude over group halo
mass is compared between fossils and controls for both simulated and
observed groups. } \label{fig6}
\end{figure}

\section{Fossil Groups: BGG Luminosity and Halo properties}

As noted earlier, fossil groups are distinguished by a large gap in
the luminosity of the two brightest members indicating the absence
of $L^*$ galaxies. If the BGGs in fossil groups are the product of
the mergers of $L^*$ galaxies, then they are expected to be
statistically brighter than the BGGs in non-fossil groups.

Fig. \ref{fig5}, shows the distribution of the r-band absolute
magnitude, $M_{r}$, for fossil and control BGGs. As expected, the
fossil BGGs are brighter than control BGGs in both the simulated and
observed groups. This is also shown in a study by \cite{diaz08}
where they use simulations to demonstrate that the fossil BGGs on
average accrete larger stellar mass as a result of earlier formation
epoch and larger frequency of mergers.

In Fig. \ref{fig6}, we compare fossil and control groups in the
plane of group mass and the BGG luminosity.  It illustrates that,
for a given r-band luminosity of the BGG, control group halos are
statistically more massive than fossil group halos.

\section{Luminosity Gap Statistics}

As noted in \S 2.3, for some of the properties studied in this
paper, there exists a probability that, random processes contribute
by part to the observed trends. In the following, we will specifically study
the contribution of random processes on halo properties of groups.


\subsection{The abundance of fossil groups}

\begin{figure}
\centering
\includegraphics[angle=0,width=75mm, clip=true]{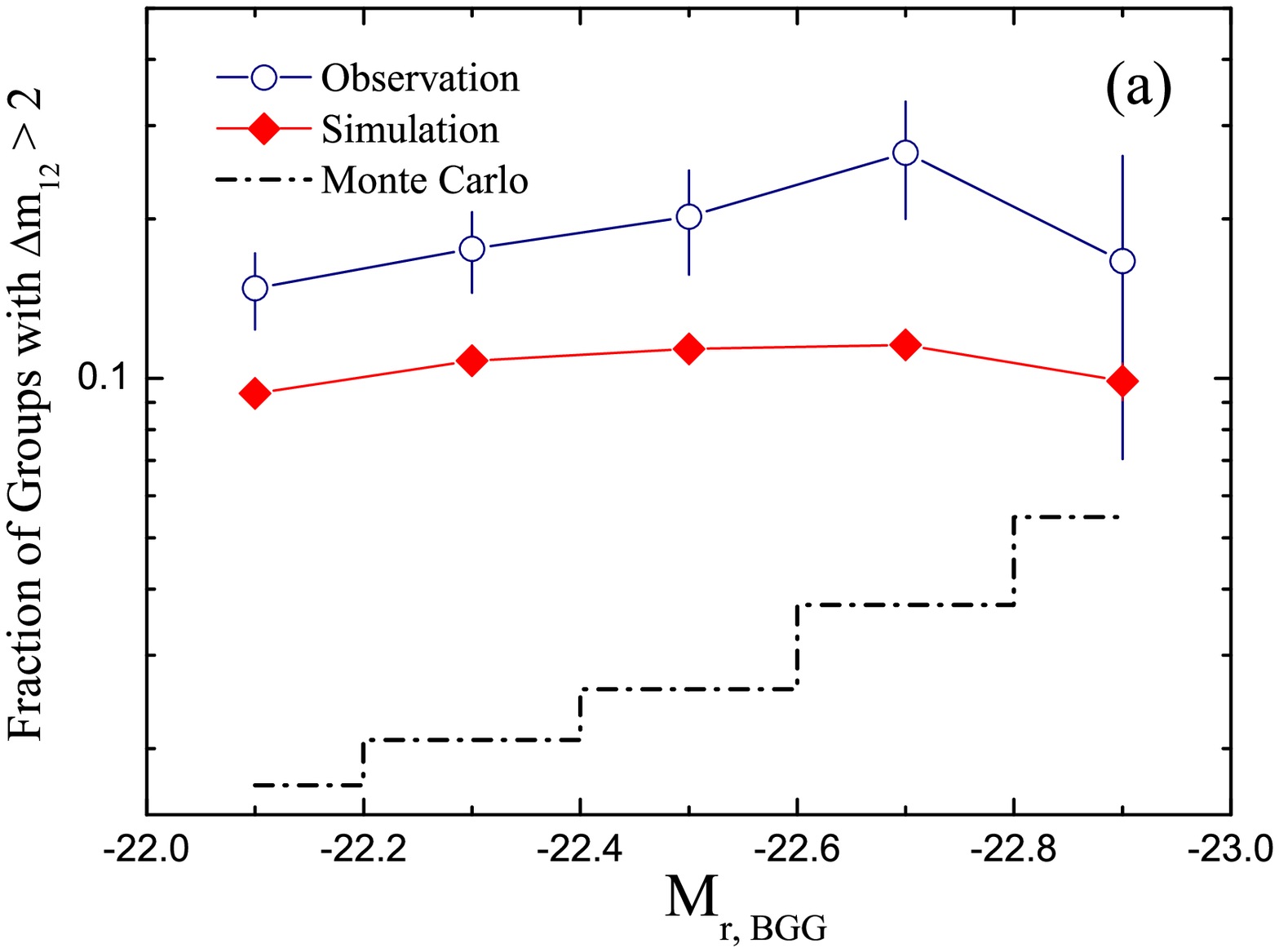}
\
\\
\
\\
\
\includegraphics[angle=0,width=75mm, clip=true]{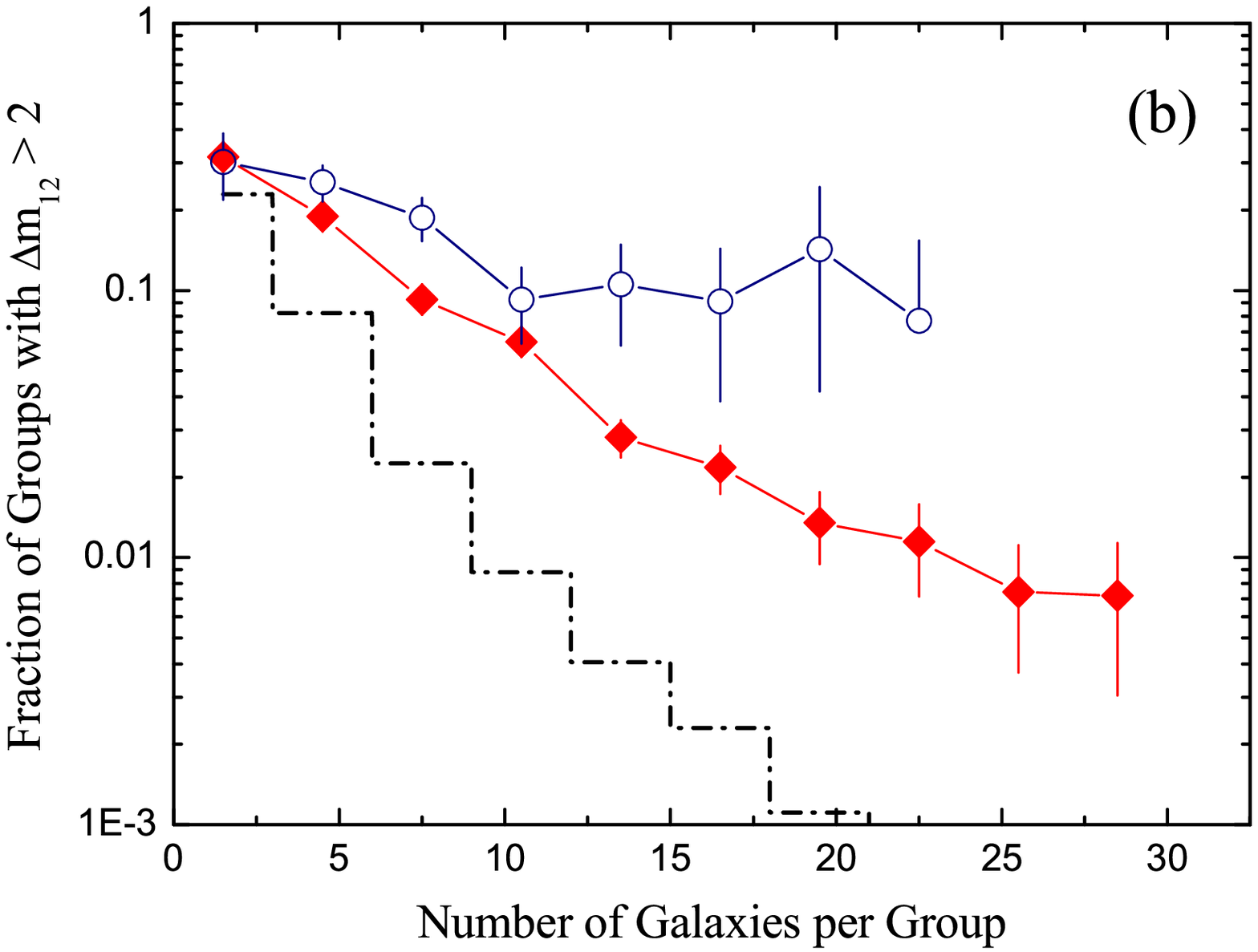}
\
\\
\
\\
\
\includegraphics[angle=0,width=75mm, clip=true]{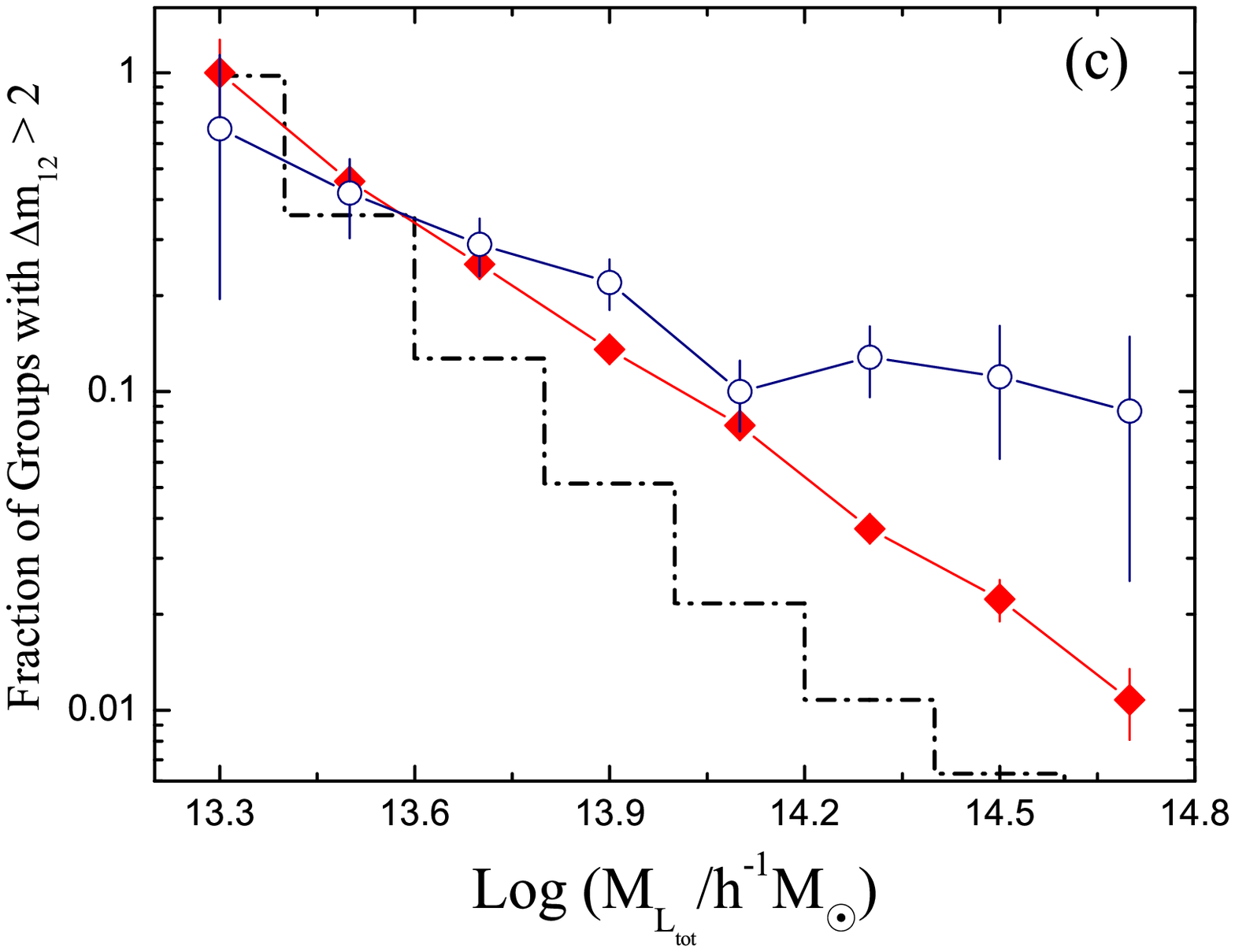}
\caption{The upper panel represents the percentage of fossil groups
in each absolute magnitude bin, for randomly generated as well as
observed and simulated groups. The percentage of fossil groups in
each richness bin is illustrated in the middle panel. The lower
panel shows the fossil percentage in each halo mass bin, for
simulated, observed and randomly generated groups. In richer and
more massive groups the evolutionary process dominate the random
processes.} \label{fig7}
\end{figure}

In Fig. \ref{fig7}-(a), we show the probability distributions of
fossil groups for a given BGG r-band luminosity, in observations as
well as in the Millennium simulation and randomly generated groups
through the Monte Carlo simulation.

As shown in Fig.\ref{fig7}-(a), within the uncertainty, the
probability of finding large luminosity gap (fossil groups) is
nearly independent of the luminosity of the brightest galaxy, in
both the observed and simulated galaxy systems. However the same
increases with the luminosity of the brightest member in Monte-Carlo
generated groups. The shape of the luminosity function is what
drives the trend seen in the Monte-Carlo generated systems. As the
probability of finding low luminosity galaxies is larger in
comparison to luminous galaxies in the Schechter function, thus, for
a fixed $M^{*}$ more luminous BGGs would result in larger luminosity
gap, statistically. The fact that such a trend is not pronounced in
the observed sample and in the cosmological simulation, supports the
argument that the large luminosity gap is unlikely to have a
statistical origin. Moreover, as shown in the Figure \ref{fig7}-(a),
the fraction of fossil groups in the observations and the simulation
is larger than the same predicted by the Monte-Carlo simulations.

Fig. \ref{fig7}-(b) shows the same as a function of group richness.
As expected, the probability of finding large luminosity gap (\dm
$\geq 2$) decreases with increasing richness. However the trend is
the fastest for the Monte-Carlo generated groups while observed
groups show a weaker dependency on richness.

In the regime of low richness, the probability distributions of the
observed, simulated and randomly generated groups coincide. As a
result, it appears that the luminosity gap in such systems is driven
by random processes, i.e., evolutionary mechanisms have played
negligible role in forming the luminosity gap.  This is supported by
the hierarchical structure formation paradigm in which the low mass
halos (poor galaxy groups) are recently formed, statistically.
\cite{dar07} concluded the same based on a comparison of a simulated
(Millennium) and Monte-Carlo generated groups. This study extends
their findings to the observed groups.

On the opposite, in richer groups, the Monte-Carlo prediction of the
large luminosity gap incidence is significantly lower than those in
cosmological simulations and the observations suggesting that the
luminosity gap in richer systems is predominantly driven by
evolutionary processes. Moreover, it is noticeable that the
cosmological simulation based on semi-analytic model of
\cite{bow06}, predicts a lower probability for fossil groups, in the
entire richness range, than the same in the observed sample.
Furthermore, choice of group finding algorithms in observations and
simulations could also contribution to this difference. The
probability distribution therefore helps to probe the accuracy of
the semi-analytic models.

In Fig. \ref{fig7}-(c) we show the fraction of fossil groups in each
mass bin. For Monte Carlo generated groups, it is prominent that,
the fraction of fossil groups in each mass bin, decrease as the halo
mass increases. As noted earlier, masses of each Monte Carlo
generated groups were assigned according to the total luminosity
method. Therefore, we expect that, both the richness of the groups
as well as the luminosity of bright members determine the halo mass.
For fossil groups (and generally for systems with large luminosity
gap), the BGG luminosity is a more defining factor because of the
presence of the large luminosity gap. Thus, fossil fraction in each
mass bin, is strongly linked to the BGG luminosity and the richness
of the group.

Comparing the probability distributions of observed, simulated and
randomly generated groups, illustrates that the three distributions
coincide in low mass end. Consequently, low mass fossil groups are
predominantly statistical i.e., evolutionary processes are not
playing a significant role in forming the luminosity gap in these
systems. On the contrary, at the high mass end, fraction of fossils
are substantially higher than what is predicted by the Monte Carlo
simulation both in the cosmological simulations and in the
observations, indicating the role of evolutionary mechanisms in
forming \dm. Interestingly semi-analytic model adopted for this
study predicts a lower fraction for fossil groups across most of the
mass/richness range. It is hard to point out at a single physical
process employed in the simulations to interpret the offset between
the fossil fraction in the observations and simulations, as this
could be a consequence of various assumptions in handling non
gravitational processes such as the starformation efficiency, AGN
feedback and cooling. These could be coupled with gravitational
processes such as the merger rate.

We propose this to be an observational constraint for the
semi-analytic models in addition to the luminosity function and the
color of galaxies.

\subsection{Luminosity Gap Probability Distribution}

\begin{figure}
\centering
\includegraphics[angle=0,width=80mm, clip=true]{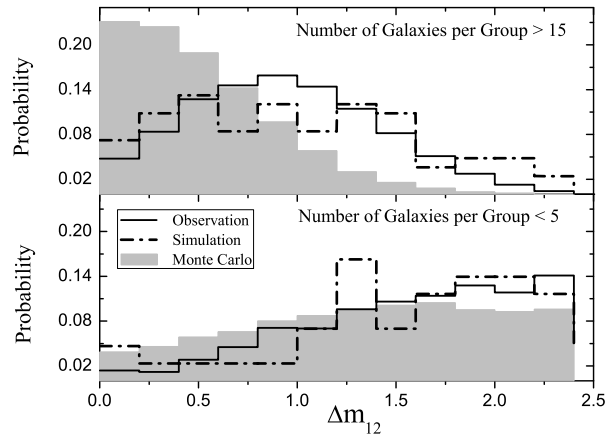}
\caption{Normalized histogram of \dm $ $ for two separate richness
regimes (see. \S 8.2).} \label{fig8}
\end{figure}

We study the probability distribution of luminosity gap. The
distributions are shown for two richness regimes, $richness \leq 5$
and $richness \geq 15$ in Fig. \ref{fig8}. It is noticeable that,
although the probability distributions in observed and simulated
groups follow a similar trend (see below), they deviate from Monte
Carlo predictions at \dm $\ge 1.2$ and \dm $\ge 0.75$ for low and
high richness respectively, such that the fraction of observed and
simulated groups with a given \dm $ $ are larger than randomly
generated groups.

We interpret this as an impact of evolution: assuming that, the
primordial distribution of \dm $ $ was what is predicted by Monte
Carlo simulation, the evolutionary mechanisms, always work in the
direction of increasing the luminosity gap.

\cite{von08} suggested an inverse process; transforming systems with
large luminosity gaps to those with low \dm. In this picture, there
is always a probability that, new galaxies fall into groups and
hence disrupt the luminosity gap. For instance, they have shown
that, luminosity gap is strongly affected by galaxy infall and hence
is a transient characteristic of galactic systems. Other processes
may also affect the luminosity gap, including mergers between galaxy
groups. However when two galaxy systems merge, they end up in
considerably more massive systems and hence with higher richness
than the progenitors. This also works in the direction of decreasing
the luminosity gap.

Fig. \ref{fig8} shows the relative efficiency of evolutionary
processes resulting in larger luminosity gap compared to inverse
processes which works in the opposite direction. The difference
between the two panels in this figure also indicates that poor
systems (low richness) on average are those that are closer to the
Monte-Carlo predictions and therefore could have recently formed and
least affected by evolutionary processes. Another possible
explanation would be that, in such systems the evolutionary
processes simply operate more slowly.

\section{New criterion for finding fossil group; \dmfo }

In a recent study based on the Millennium simulation by
\cite{dar10}, a new indicator based on gap magnitude between first
and forth brightest galaxies within half of the virial radius of the
group {\dmtfo} is presented to be a more efficient probe of
identifying early-formed halos than conventional definition of
fossil galaxy group, i.e. {\dmt}.

\cite{dar10} find that the mass assembly histories of the halos
identify by the two methods, on average, are similar. About $90\%$
of fossil groups which were identified according to {\dm} and
{\dmfo} criteria in earlier epochs become non-fossils after $\sim4
Gyr$ and the fossils phase itself lasts $\sim1 Gyr$. The main
difference between the two methods seem to be in the efficiency of
finding early formed halos (\citealt{dar10}).

In order to evaluate the abundance of fossil groups according to
these definitions, we compared the luminosity gap distribution based
on simulation and observation data (our catalog).

To be able to estimate the luminosity gap parameter {\dmfo}, groups
with the following properties are selected. The extracted groups
have at least four members within the half of the virial radius of
the galaxy group and the brightest and fourth brightest galaxy
should be at least as luminous as $-22$ and $-19$, respectively.
Applying above criteria, $163$ groups were identified in our
catalog, out of which $41$ groups meet the {\dmtfo} criterion.

Among these observed fossil groups ({\dmtfo}), $12\%$ satisfy the
conventional definition of fossil ({\dmt}). Conversely, $56\%$ of
{\dmt} satisfy the {\dmtfo} criterion. Hence, a large proportion of
the population of groups identified with {\dmtfo} criterion are not
in common with those identified by {\dmt}. This is consistent with
the findings of \cite{dar10} based on simulations.

We applied the same criteria to extract galaxy group in simulation
data. Out of $16688$ groups, $1993$ and $2973$ groups satisfy the
{\dmt} and {\dmtfo} groups respectively.

\begin{figure}
\centering
\includegraphics[angle=0,width=80mm, clip=true]{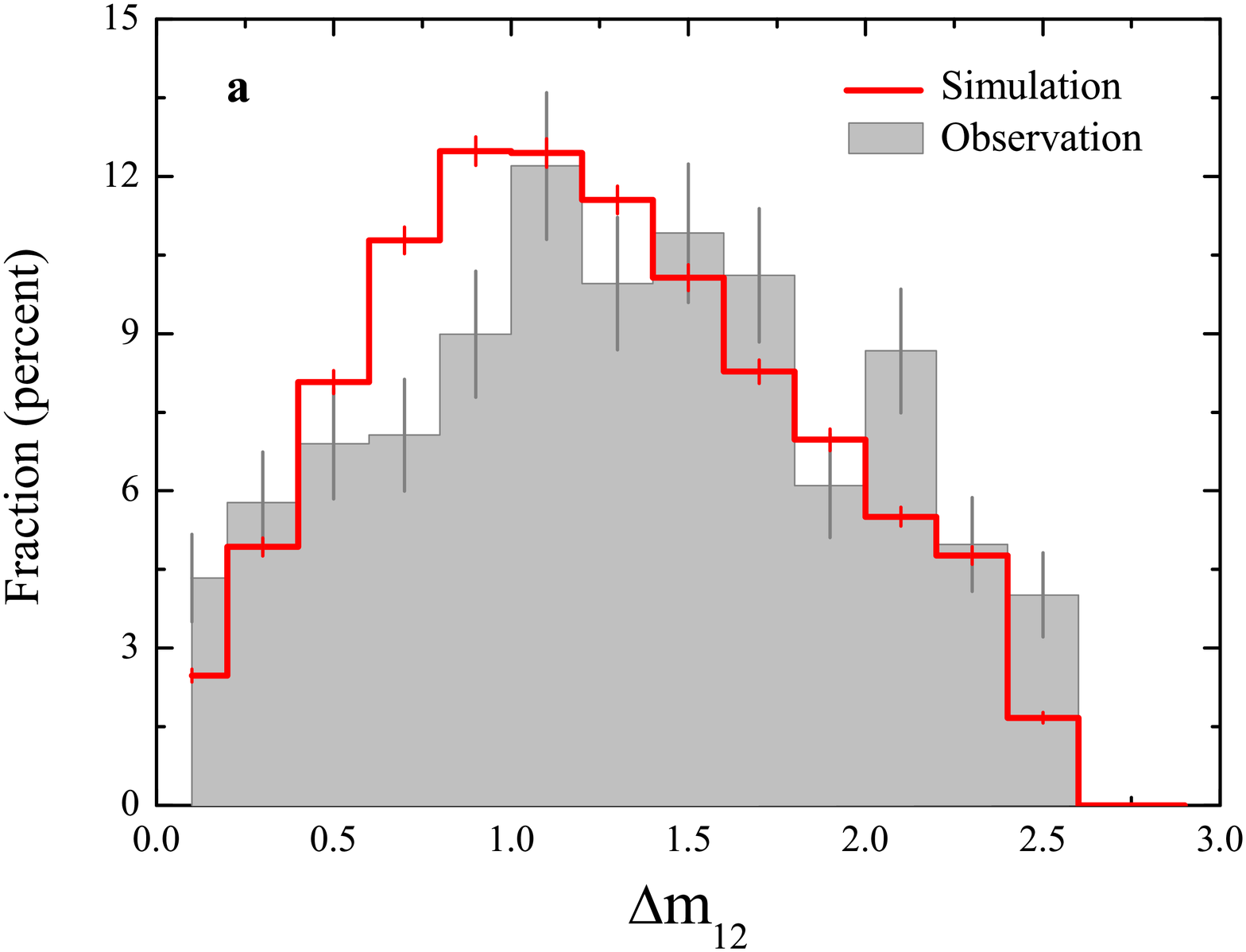}
\includegraphics[angle=0,width=80mm, clip=true]{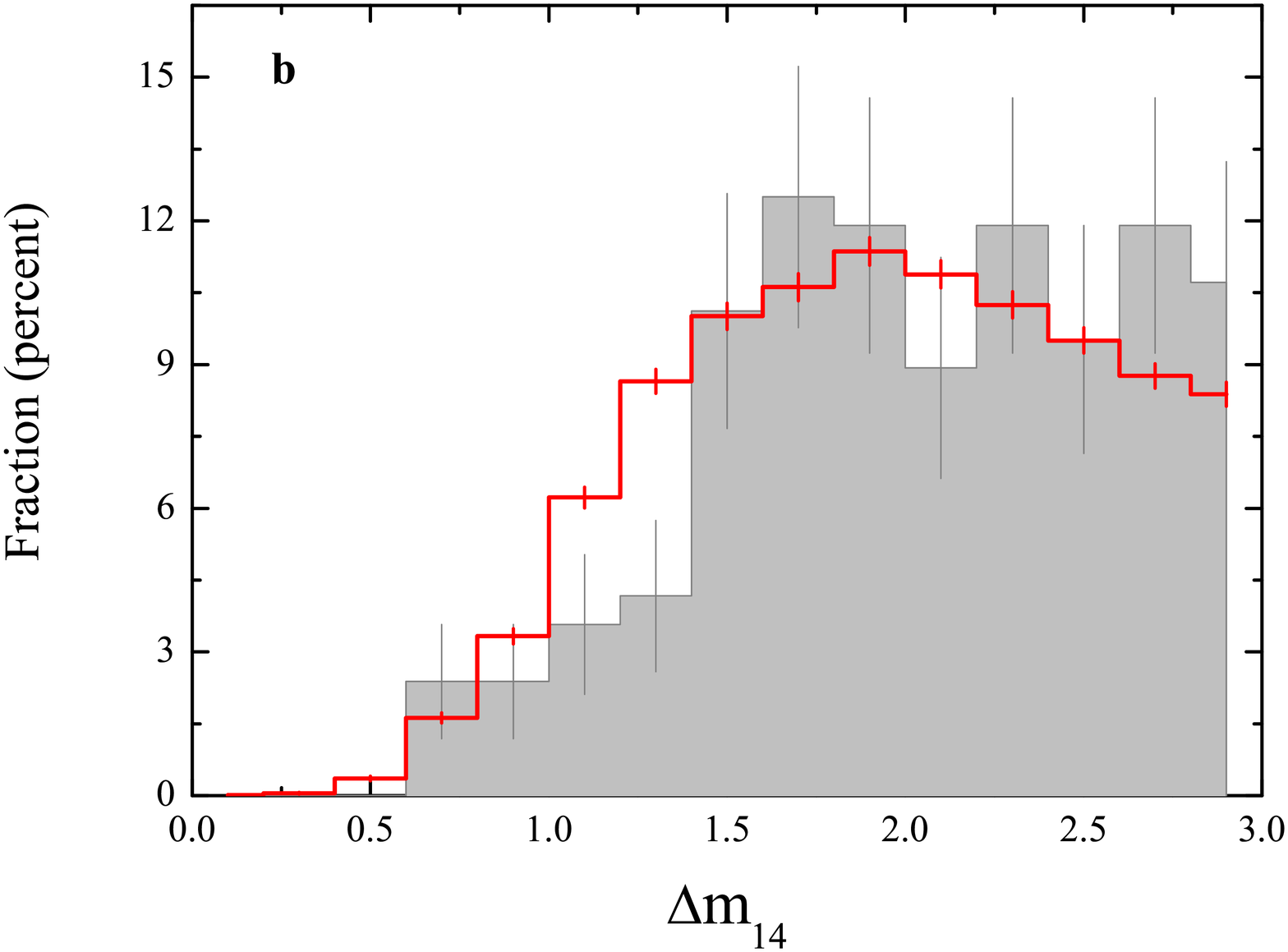}
\caption{Distribution of \dm (\textbf{a}) and \dmfo (\textbf{b}) for
observational and simulation samples.} \label{fig9}
\end{figure}

In Fig.\ref{fig9}, we plot the R-band luminosity gap distribution
{\dm} and {\dmfo} from our observations and simulations. Results
show that the luminosity gap from simulation is in a fair agreement
with observations and both present a similar distributions for the
{\dm} and {\dmfo} luminosity gap. The fraction of groups with {\dmt}
in observation and simulation are $17\%$ and $12\%$, respectively
while these values for {\dmtfo} indicator are $25\%$ and $18\%$,
respectively.

The probability of finding fossil system according to these
definitions in current analysis is different to the statistics
reported by \cite{dar10} which uses the group catalog of
\cite{yang07} based on $SDSS$ $DR4$. For instance \cite{dar10}
reports $2.0\%$ and $2.1\%$ in observation and simulation,
respectively when using {\dmt}. For {\dmtfo}, they find $6.2\%$ and
$5.1\%$, respectively. The difference is due to the completeness
considerations in sample selection adapted by us.

The most important difference between our group catalog and that of
\cite{yang07} is the addition constrain we have imposed on BGG
luminosity and the luminosity gap (see. \S 4). Furthermore their
groups are defined as systems which at least four members while we
require the group to have two spectroscopic member within half of
the virial radius.

\begin{figure}
\centering
\includegraphics[angle=0,width=80mm, clip=true]{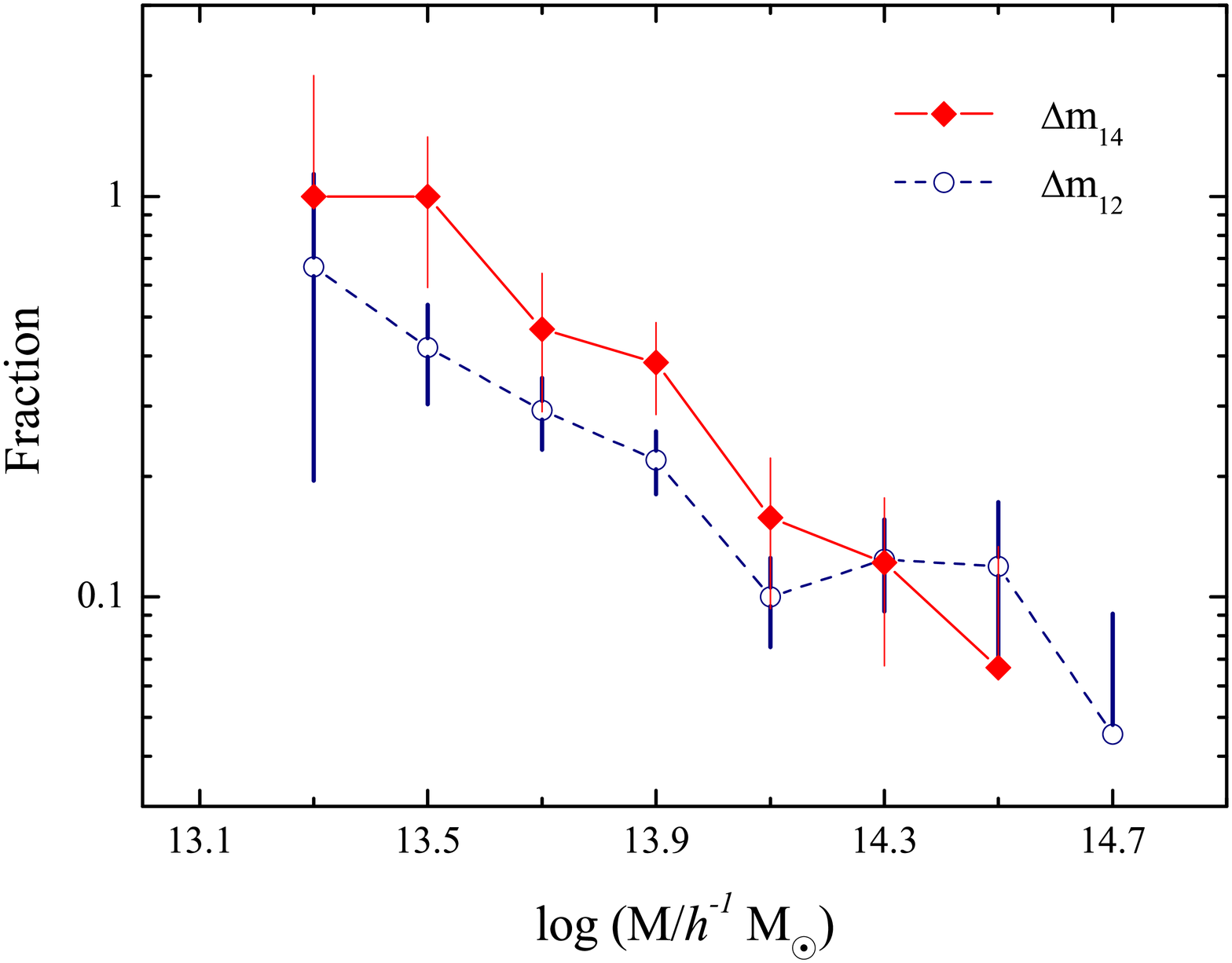}
\caption{Fraction of fossil groups according to both definitions
\dmt $ $ and \dmtfo, as a function of halo mass, in observational
sample.} \label{fig10}
\end{figure}

In \S 8.1, it was found that the probability of finding early-formed
system, i.e. fossil groups  {\dmt}, is higher in low mass systems.
In Fig.\ref{fig10}, the abundance of fossil groups base on
conventional definition {\dmt} and the new indicator {\dmtfo} are
given as function of halo mass. This shows that, in comparison to
conventional fossils ({\dmt}), the fossil groups based on {\dmtfo}
are more abundant in low mass range $log (M/h^{-1} M_\odot) < 14.3$.
Above this mass limit, the statistics is poor and therefore hard to
interpret.

\section{Discussion and Conclusion}

In this study we extend the earlier observational studies of
luminosity gap to low mass regime and provide observational
constrains for semi-analytic models based on this observable. We
extract 620 groups in SDSS DR7 utilizing mFoF algorithm and measure
the luminosity gap between the two brightest galaxies in the group,
{\dm}. This results in 109 groups with {\dmt}, within half a virial
radius, known as optical fossils groups. In addition, \cite{bow06}
semi-analytic model employed on Millennium simulation was used to
select 16688 groups with 1993 fossil groups in the same manner as
applied to the observational data. A Monte carlo simulation was also
performed to generate a large sample of luminosity gap statistics
purely drawn at random from Schechter luminosity function to
investigate the importance of the random processes.

We show that fossil BGGs both in observations and simulations are on
average brighter than those in control galaxy groups.

Given a relatively poor constrain on galaxy velocity dispersion in
groups (e.g. dynamical mass estimation) we adapted the total
luminosity method for the estimation of the total gravitational mass
of the groups based on empirical relations. We show fossil groups
halos are found to be statistically less massive than those of the
control sample for a given BGG luminosity.

We confirm that the luminosity gap in systems with low richness is
predominantly driven by random processes while evolutionary
mechanisms are responsible for large luminosity gap in rich groups.

We find indications that the fraction of fossil groups based on
{\dmtfo} criterion is higher compared to the same when adopting the
conventional {\dmt} criterion. This is valid mostly to the low mass
end.

\begin{figure}
\centering {\includegraphics[angle=0,width=80mm, clip=true]{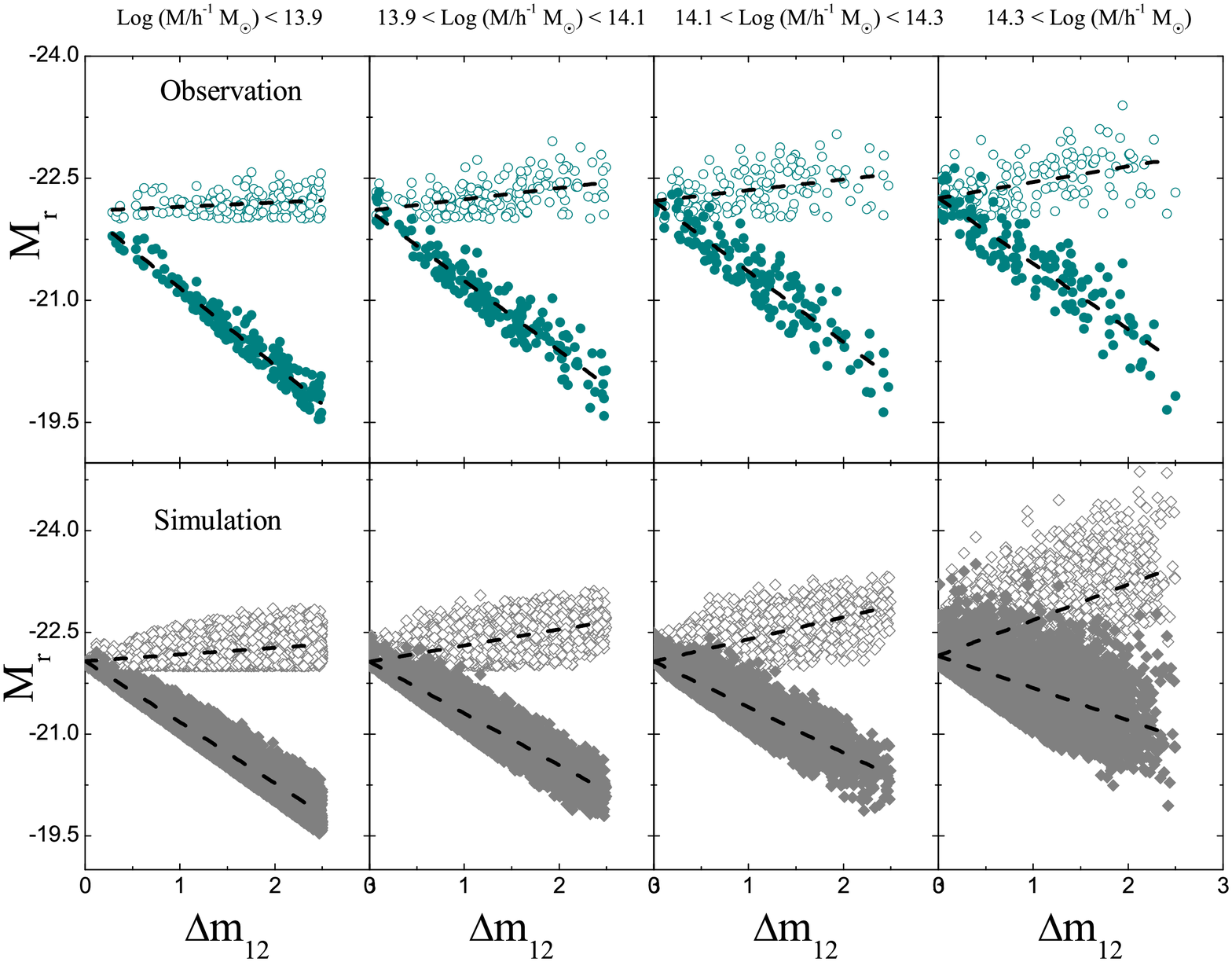}
\vspace{0.2in}}
{\includegraphics[angle=0,width=80mm,clip=true]{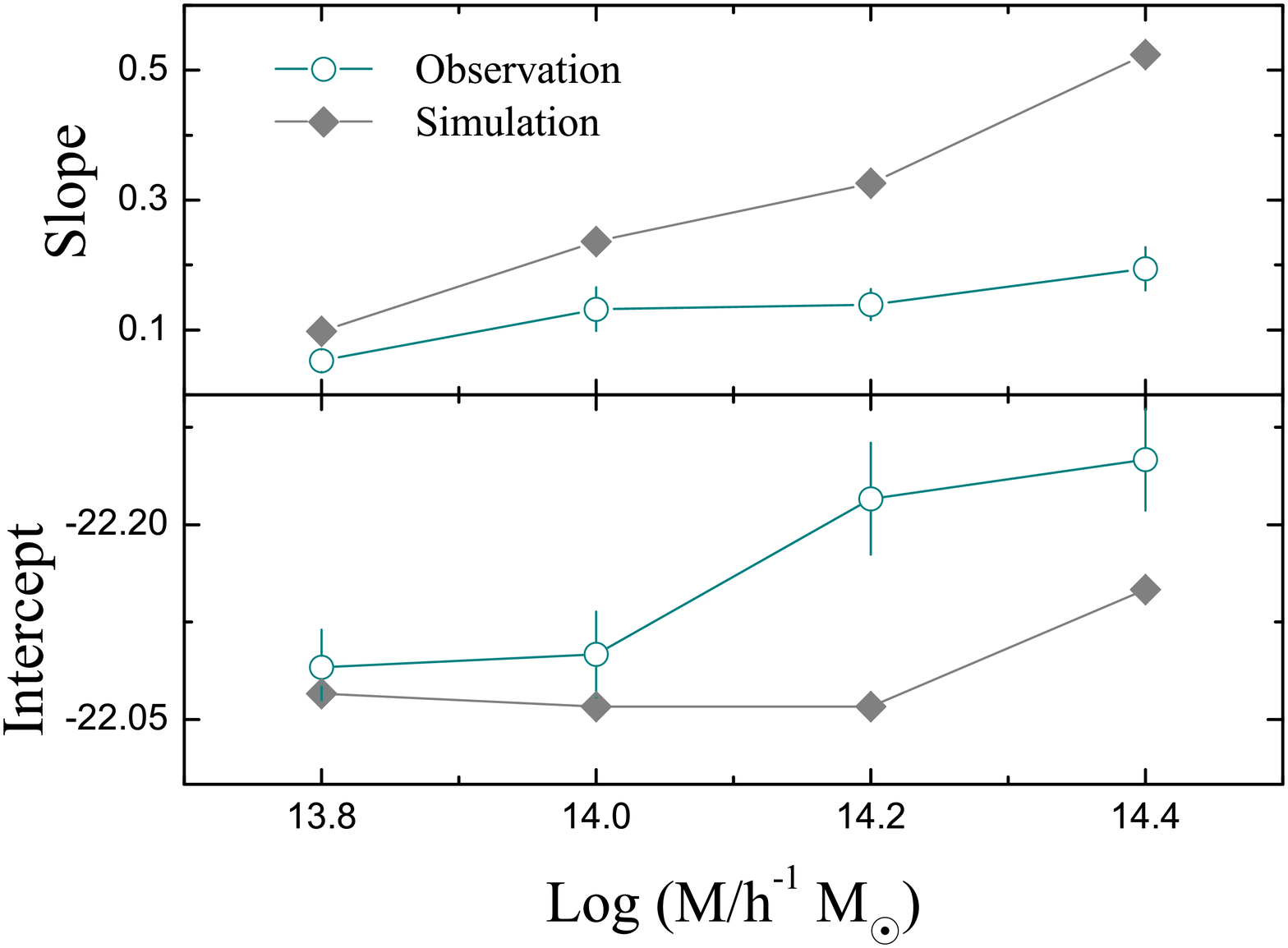}}
\caption{Upper Figure shows the distribution of r-band absolute
magnitude for the first (open circles) and second (filled circles)
brightest group member as a function of $\triangle m_{12}$ for
different halo masses. Figure at the bottom shows, the evolution of
slopes and intercepts of brightest group member as a function of
halo mass.} \label{fig11}
\end{figure}

Following a recent study by \cite{smi09}, base on a sample of
massive galaxy clusters $(\sim 10^{15} M_{\odot})$, we propose a
test based on the luminosity gap in galaxy group, the butterfly
diagram $M_{r}$ vs. \dm, which is found to be able to discriminate
between different galaxy formation models in groups and clusters
when a large sample of groups with known luminosity gap is
available. Extending their study to low mass groups, we compare 620
observed groups and 16688 simulated groups in the plane of $M_{r}$
vs. \dm$ $, Fig. \ref{fig11}. The absolute magnitude of the
brightest and second brightest galaxy within the group are plotted
as a function of \dm $ $ for different halo masses. A linear
regression is also presented for both the brightest and the second
brightest galaxies as a function of \dm.  In addition, Fig.
\ref{fig11} shows the variation of the slope and the intercept of
the linear fit as a function of the halo mass for BGGs.

We find that the slope of the BGG brightness in \cite{bow06}
simulation is steeper than the observed BGGs, particulary in high
mass regime. This implies that the AGN feedback in BGG is too weak
in \cite{bow06} semi-analytic model \citep{smi09}. Further
improvements have been made in the semi-analytic model recently
(Bower et al 2008) driven by constraints from the IGM properties of
the galaxy systems, which also required a re-tuning of the galaxy
formation parameters in their earlier model to be able to
reconstruct the galaxy luminosity function in the local universe.

The discussion on the success of their changes in the context of the
luminosity gap requires a direct comparison of the two semi-analytic
models \cite{bow06,bow08}. We find that the BGG brightness of
observed groups, increases with halo mass. In contrast this trend is
not clear in simulated groups. Furthermore, in high mass bin, the
absolute magnitude of simulated BGG span $\sim 2.5$ mag, in contrast
to the observed range of $\sim 1$ mag. \cite{smi09} argue that the
large spread in the absolute magnitude of simulated BGGs is driven
by higher efficiency of the conversion of the cold gas into stars.
We note that in low mass bin, where the random processes are
dominated, the spread in the luminosity of the observed and
simulated BGGs are nearly the same ($\sim 0.5$   $mag$).

The correlation between the slope of the line fitted to BGG in Fig.
\ref{fig11} and halo mass points to an increasing contribution of
the evolutionary mechanisms in forming the luminosity gap in more
massive halos. Giving the considerable space density of large
luminosity gap groups (e.g. \citealt{vikh99}, \citealt{J03},
\citealt{von07}, \citealt{d05}, \citealt{von08}, \citealt{dar07}),
it seems that the butterfly diagram is a simple way of testing the
accuracy of the semi-analytic models.

\section*{Acknowledgments}

Funding for the $SDSS$ and $SDSS-II$ has been provided by the Alfred
P. Sloan Foundation, the Participating Institutions, the National
Science Foundation, the U.S. Department of Energy, the National
Aeronautics and Space Administration, the Japanese Monbukagakusho,
the Max Planck Society, and the Higher Education Funding Council for
England. The SDSS Web Site is $http://www.sdss.org/$. The SDSS is
managed by the Astrophysical Research Consortium for the
Participating Institutions. The Participating Institutions are the
American Museum of Natural History, Astrophysical Institute Potsdam,
University of Basel, University of Cambridge, Case Western Reserve
University, University of Chicago, Drexel University, Fermilab, the
Institute for Advanced Study, the Japan Participation Group, Johns
Hopkins University, the Joint Institute for Nuclear Astrophysics,
the Kavli Institute for Particle Astrophysics and Cosmology, the
Korean Scientist Group, the Chinese Academy of Sciences (LAMOST),
Los Alamos National Laboratory, the Max-Planck-Institute for
Astronomy (MPIA), the Max-Planck-Institute for Astrophysics (MPA),
New Mexico State University, Ohio State University, University of
Pittsburgh, University of Portsmouth, Princeton University, the
United States Naval Observatory, and the University of Washington.

The Millennium simulation used in this paper was carried out by the
Virgo Supercomputing Consortium at the Computing Centre of the
Max-Planck Society in Garching. The semianalytic galaxy catalogue is
publicly available at $http://www.mpagarching.
mpg.de/galform/agnpaper$.

\clearpage











\begin{thebibliography}{}

\bibitem[Abazajian et al.(2009)]{aba09}
Abazajian K., et al., 2009, ApJS, 182, 543

\bibitem[Abell (1958)]{ab58} Abell, G.O., 1958, ApJS, 3,211

\bibitem[Barnes (1989)]{ba89} Barnes, J.E., 1989, Nature, 338,123

\bibitem[Beers et al.(1990)]{bee90}Beers, T.C., Flynn, K., Gebhardt, K., 1990, AJ, 100, 32

\bibitem[Berlind et al.(2006)]{ber06} Berlind, A. A., et al. 2006, ApJ, 167, 1

\bibitem[Blanton et al.(2003a)]{bla03a} Blanton, M. R., et al. 2003a, AJ, 125, 2276

\bibitem[Blanton et al.(2003b)]{bla03b} Blanton, M. R., et al. 2003b, AJ, 125, 2348

\bibitem[Bower et al.(1992)]{bow92}Bower,R., Lucey, J. R., Ellis, R. S., 1992, MNRAS, 254, 601

\bibitem[Bower et al.(2006)]{bow06} Bower, R. G.; Benson, A. J.; Malbon, R.; Helly, J. C.;Frenk
 ,C. S.; Baugh, C. M.; Cole, S.; Lacei, C. G., 2006, MNRAS, 370,645

\bibitem[Bower et al.(2008)]{bow08} Bower, R. G., McCarthy, I. G., \& Benson, A. J. 2008, MNRAS,
390, 1399

\bibitem[Brough et al.(2008)]{bro08}
Brough, S., Couch, W. J., Collins, C. A., Jarrett, T., Burke, D. J.,
\& Mann, R. G. 2008, MNRAS, 385, 103

\bibitem[Colless et al.(2001)]{col01} Colless, M. et al., 2001,
MNRAS, 328 ,1039

\bibitem[Croton et al.(2006)]{cro06} Croton, D.J., Springel, V., White, S.D.M., De Lucia, G.; Frenk, C.
S.; Gao, L.; Jenkins, A., Kauffmann, G., Navarro, J. F., \& Yoshida,
N., 2006, MNRAS, 365, 11

\bibitem[Dariush et al.(2007)]{dar07} Dariush, A., Khosroshahi,
H. G., Ponman, T. J., Pearce, F.,Raychaudhury, S., \& Hartley, W.
2007, MNRAS, 382, 433

\bibitem[Dariush et al.(2010)]{dar10} Dariush, A., Raychaudhury, S., Ponman, T. J., Khosroshahi, H. G.,
Benson A. J., Bower R. G., \& Pearce, F., 2010, MNRAS, 405, 1873

\bibitem[Diaz-Gimenez et al.(2008)]{diaz08} Diaz-Gimenez E., Muriel H.,
Mendes de Oliveira C., 2008, A\&A, 490,965


\bibitem[D'Onghia et al.(2005)]{d05} D'Onghia E., Sommer-Larsen J., Romeo A. D., Burkert A., Pedersen K.,
Portinari L., Rasmussen J., 2005, ApJ, 630, 109

\bibitem[Einasto et al.(2005)]{ein05}
Einasto, J., Tago E., Einasto, M., Saar, E.: 2005a, In "Nearby
Large-Scale Structures and the Zone of Avoidance", eds. A.P.
Fairall, P. Woudt, ASP Conf. Series, 329, 27

\bibitem[Ellis \& O'Sullivan (2006)]{ellis06}
Ellis S. C., O'Sullivan E., 2006, MNRAS, 367, 627

\bibitem[Girardi et al.(1998)]{gir98} Girardi M., Giuricin G., Mardirossian F., Mezzetti M., Boschin W.,
1998, ApJ, 505, 74

\bibitem[Gladders et al.(2000)]{gla00}Gladders, M. D., Yee, H. K. C., 2000, AJ, 120, 2148

\bibitem[Goto et al.(2005)]{got05} Goto, T., et al. 2005, MNRAS, 359, 1415

\bibitem[Huchra \& Geller (1982)]{huch82}
Huchra, J. P., \& Geller, M. J. 1982, ApJ, 257, 423

\bibitem[Jones et al.(2003)]{J03} Jones, L.R, Ponman, T.J., Horton, A., Babul, A., Ebeling, H.
Burke, D.J.,  Forbes, D.A., 2003, MNRAS, 343, 627

\bibitem[Khosroshahi et al.(2006)]{kh06b}
Khosroshahi H. G., Maughan B. J., Ponman T. J., Jones L. R., 2006, MNRAS, 369, 1211

\bibitem[Khosroshahi, Ponman \& Jones (2006)]{kh06}
Khosroshahi H. G., Ponman T. J., Jones L. R., 2006, MNRAS, 372, L68

\bibitem[Khosroshahi et al.(2007)]{kh07} Khosroshahi, H.G.,
 Ponman, T.J. \& Jones, L.R., 2007, MNRAS, 377, 595

\bibitem[Kodama et al.(1998)]{koda98} Kodama, T., \& Arimoto, N. 1998, MNRAS, 300, 193

\bibitem[Lin et al.(2004)]{lin04} Lin, Y. T., Mohr, J. J.,  Stanford, S. A. 2004, ApJ, 610, 745

\bibitem[Merchan et al.(2005)]{mer05} Merchan M., Zandivarez A., 2005, ApJ, 630, 759

\bibitem[Miller et al.(2005)]{mil05} Miller, C. J., et al. 2005, AJ, 130, 968

\bibitem[Milosavljevic et al.(2006)]{milo06} Milosavljevic, M., Miller C. J., Furlanetto R., Cooray A.,
2006, ApJ, 637, 9

\bibitem[Oyaizu et al.(2008)]{oya08} Oyaizu, H., et al. 2008, ApJ, 2008,
674, 768

\bibitem[Paredes et al.(1995)]{par95} Paredes S., Jones
B. J. T., Martinez V. J., 1995, MNRAS, 276, 1116

\bibitem[Peebles et al.(2002)]{peb02}Peebles, P. J. E. 2002, in ASP Conf. Ser. 283, A New Era in
Cosmology, ed. N. Metcalfe \& T. Shanks (San Francisco: ASP), 351

\bibitem[Petrosian (1976)]{pet76} Petrosian, V. 1976, ApJ, 210, 53

\bibitem[Popesso et al.(2005)]{pop05} Popesso, P., Biviano, A., Bhringer, H., Romaniello, M.,  Voges, W. 2005,
A\&A, 433, 431

\bibitem[Ponman et al.(1994)]{p94} Ponman, T.J., Allan, D.J., Jones, L.R., Merrifield, M.  MacHardy,
I.M., 1994 Nature, 369, 462

\bibitem[Santos et al.(2007)]{san07} Santos,W.A., Mendes de Oliveira, C., \& Sodr\'{e}, L., 2007, AJ, 134,
1551

\bibitem[Schechter (1976)]{sch76} Schechter P., 1976, ApJ, 203, 297

\bibitem[Schlegel et al.(1998)]{schl98} Schlegel, D. J., Finkbeiner, D. P., \& Davis, M. 1998, ApJ, 500, 525

 \bibitem[Smith et al.(2010)]{smi09}Smith,
 G.P., Khosroshahi, H.G.,  Dariush, A.,
 Sanderson, A.J.R., Ponman, T.J., Stott, J.P., Haines, C.P., Egami, E., Stark, D.P., 2010,
 ,MNRAS,in press,arXive:1007.2196S

\bibitem[Spergel et al.(2003)]{spe03} Spergel, D. N. et al., 2003, ApJS, 148, 175

\bibitem[Springel et al.(2001)]{spr01} Springel, V., White, S.D.M.,
Tormen, G., \& Kauffmann, G., 2001, MNRAS, 328, 726

\bibitem[Springel et al.(2005)]{spr05} Springel, V. et al.,
2005, Nature, 435, 629

\bibitem[Strauss et al.(2002)]{str02} Strauss, M. A. et al. 2002, AJ, 124, 1810

\bibitem[Struble et al.(1987)]{stru87} Struble M.F., Rood H.J. 1987, ApJS, 63, 555

\bibitem[Tago et al.(2008)]{tag08} Tago, E. et al. 2008, A\&A, 479, 927-937

\bibitem[van den Bosch (2007)]{von07} van den Bosch F. C. et al. 2007,
MNRAS, 376, 841

\bibitem[Vikhlinin et al.(1999)]{vikh99} Vikhlinin, A.,McNamara, B.R., Hornstrup, A., Quintana, H., Forman,
W., Jones, C.  Way, M., 1999, ApJ, 520, 1

\bibitem[Voges et al.(1999)]{vog99}
Voges, W., et al. 1999, A\&A, 349, 389

\bibitem[von Benda-Beckmann et al.(2008)]{von08}von Benda-Beckmann, A. M., D'Onghia, E., Gottlber, S., Hoeft, M.,
Khalatyan, A., Klypin, A., \& Mller, V., 2008, MNRAS, 386, 2345

\bibitem[von der Linden et al.(2007)]{vonder07}von der Linden A., Best P. N., Kauffmann G., White S. D. M., 2007,
MNRAS, 379, 867

\bibitem[Yang et al.(2005)]{yang05}
Yang X., Mo H.J., van den Bosch F.C., Jing Y.P., 2005a,
MNRAS,356,1293

\bibitem[Yang et al.(2007)]{yang07}
Yang X. H., Mo H. J., van den Bosch F. C., Pasquali A., Li Ch.,
Barden M., 2007, ApJ, 671, 153

\bibitem[Yoon et al.(2008)]{yoon08} Yoon J. H., Schawinski K., Sheen Y., Ree C. H., Yi S. K., 2008, ApJS, 176, 414

\bibitem[Zandivarez et al.(2006)]{zan06} Zandivarez A.,Martinez H.J., Merchan M., 2006, ApJ, 650, 137

\end{thebibliography}
\end{document}